\documentclass[a4paper]{article}

\usepackage{icrc2013}
\usepackage[english]{babel}

\title{On the GCR intensity and the inversion of the heliospheric magnetic field during\- the periods of the high solar activity }

\shorttitle{GCR intensity and the inversion of the HMF}

\authors{
Krainev M.B., Kalinin M.S.
}

\afiliations{
Lebedev Physical Institute, Russian Academy of Sciences, Moscow, Russia
}

\email{mkrainev46@mail.ru}

\abstract{We consider the long-term behavior of the solar and heliospheric parameters and the GCR intensity in the periods of high solar activity and the inversions of heliospheric magnetic field (HMF). The classification of the HMF polarity structures and the meaning of the HMF inversion are discussed. The procedure is considered how to use the known HMF polarity distribution for the GCR intensity modeling during the periods of high solar activity. We also briefly discuss the development and the nearest future of the sunspot activity and the GCR intensity in the current unusual solar cycle 24.
}


\begin{document}
\maketitle

%Begin a section.
\section{Introduction}
There are several interesting features in the solar activity, heliospheric characteristics and the GCR intensity in the maximum phase of the solar cycle. First, the sunspot area and the HMF strength are at their highest levels during these periods and both demonstrate the two--peak structure with the Gnevyshev Gap between the peaks (see \cite{Storini_Felici_NuovoCmento_17C_697_1994,Bazilevskaya_etal_SP_197_157-174_2000} and references therein). Second, the inversion of the high--latitude solar magnetic fields (SMF) occurs in this phase \cite{Howard_Solar Physics_38_283_1974,Makarov_Makarova_SP_163_267_1996} and it changes the distribution of the HMF polarity in the heliosphere. Third, as the GCR intensity in general anticorrelates with the sunspot area and HMF strength, this intensity is rather low in these periods and it demonstrates the two--gap structure corresponding to the the two--peak structure of the sunspot area and HMF strength with somewhat different behavior for the low and high energy particles (the energy hysteresis, see \cite{Moraal_ICRC14_11_3896-3906_1975,Krainev_Bazilevskaya_ASR_35_2124-2128_2005} and references therein). Note that these phenomena can be very specific in the maximum phase of the current unusual solar cycle (SC) 24.

The LPI cosmic--ray group has been studying the complex of these phenomena for more than 40 years, traditionally connecting the GCR intensity behavior with the SMF inversion (see \cite
{Charakhchyan_etal_ICRC13_2_1159-1164_1973,Vernov_etal_ICRC14_3_1015_1975,Krainev_etal_ICRC19_4_481-484_1985,
Bazilevskaya_etal_ASR_9_227-231_1995,Krai61_ICRC26_7_155-158_1999,Krainev_Bazilevskaya_ASR_35_2124-2128_2005} among others). It is suffice to mention that the first papers on the possible SMF influence on the GCR intensity were \cite{Charakhchyan_etal_ICRC13_2_1159-1164_1973} and \cite{Vernov_etal_ICRC14_3_1015_1975} dealing with the energy hysteresis.

In this paper we first reanimate an old scenario of ours on the long-term behavior of the solar and heliospheric parameters and the GCR intensity and correlation between them in the periods of high solar activity and the inversions of the large--scale SMF. Then we discuss the classification of the HMF polarity structures in order to clarify the meaning of the HMF inversion and consider how to model the GCR behavior in these periods. Finally we discuss the development of the current SC 24 and the maximum  sunspot area and minimum GCR intensity which could be expected in the near future.

\section{The SMF, HMF and GCR intensity during the periods of high solar activity}
\begin{figure}[h]
  \centering
  \includegraphics[width=0.5\textwidth]{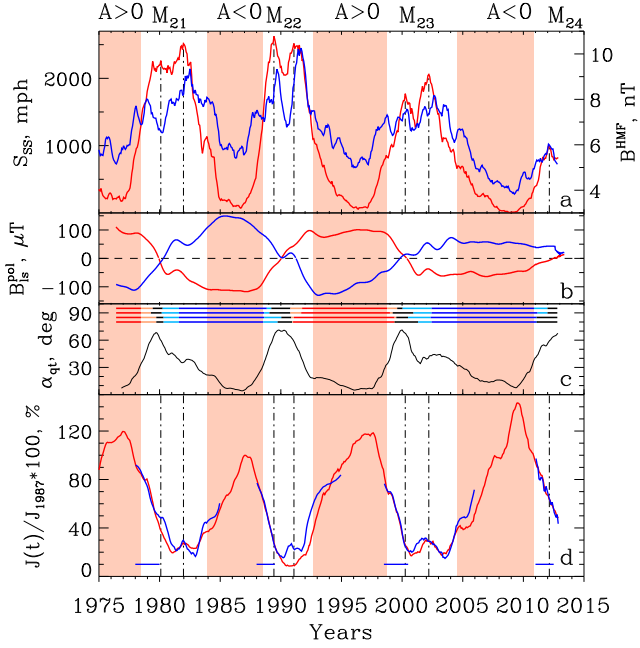}
  \caption{The solar activity, heliospheric parameters and GCR intensity in 1975--2013. The periods of low solar activity are shaded and the HMF polarity $A$ and the moments of the sunspot maxima are indicated above the panels.
  All data are yearly smoothed. In the panels:
  (a) the total sunspot area $S_{ss}$ (red) and the HMF strength near the Earth $B^{HMF}$ (blue). The moments of two (for each cycle except SC 24) Gnevyshev gaps in $S_{ss}$ are shown by the vertical dash--dot lines in the panels (a) and (d); (b) the line--of--sign components of the high--latitude photospheric SMF in the N (red) and S (blue) hemispheres; (c) the HCS quasi--tilt (black) and the classification of the periods with respect to the HMF polarity states (red and blue sections of four horizontal lines in the upper part of the panel, see the next section); (d) the low energy GCR intensity normalized to 100\% in 1987 (red) and the high energy intensity (blue) for the periods of high solar activity regressed to that for the low energies by the the linear regression in the periods indicated by the horizontal blue lines near the time axis.
  }
  \label{fig1}
 \end{figure}

In Fig. \ref{fig1} the time profiles of all related characteristics are shown for the last 40 years.
The data on the sunspot area $S_{ss}$ \cite{Sss_Site}, the HMF strength near the Earth $B^{HMF}$ \cite{OMNI_Site} and the SMF characteristics (the quasi--tilt $\alpha_{qt}$ of the heliospheric current sheet (HCS), the high--latitude SMF $B^{pol}_{ls}$ and the sets of the spherical harmonic coefficients $\left\{g_{l,m},h_{l,m}\right\}$ for the Wilcox Solar Observatory (WSO) model) \cite{WSO_Site} are used. As a low energy GCR data we use the results of our stratospheric regular balloon monitoring (RBM):  the difference between the count rates of the omnidirectional counter in the Pfotzer maximum in Murmansk $N_{RBM}^{Mu}$ (cutoff rigidity $R_c\approx 0.6$ GV) and the same characteristic in Moscow $N_{RBM}^{Mo}$ ($R_c\approx 2.4$ GV) \cite{Bazilevskaya_Svirzhevskaya_SSR_85_431-521_1998,Stozhkov_etal_Preprint_LPI_14_2007}. As a high energy GCR intensity  we use the neutron monitor data (Moscow, the effective energy $T_{eff}\approx 15$ GeV, \cite{NM_Moscow}).

 In Fig. \ref{fig1} one can easily see for SC 21--23 the double-peak structure of the sunspot maximum phase both in $S_{ss}$ and $B^{HMF}$  and corresponding double-gap signature in the GCR intensity. This phenomenon was called Gnevyshev Gap in \cite{Storini_Felici_NuovoCmento_17C_697_1994} and was extensively studied (see \cite{Bazilevskaya_etal_SP_197_157-174_2000,Krainev_Bazilevskaya_ASR_35_2124-2128_2005} and references therein). In \cite{Bazilevskaya_etal_SP_197_157-174_2000} we came to the conclusion that the double peak structures during solar maxima are due to the superposition of the 11-year cycle and quasi-biennial oscillations. However, note that the Gnevyshev gaps usually coincide or occur just after the inversions of the high--latitude SMF shown in Fig. \ref{fig1} (b) so there could be some physical connection between these two phenomena. In Fig. \ref{fig1} (c) the quasi--tilt $\alpha_{CS}$ (classic) is shown as well as the four horizontal lines illustrating the suggested classification of the periods with respect to the HMF polarity (see next section).

Another GCR effect specific for the maximum phase of solar cycle and also seen in Fig. \ref{fig1} (d) is the energy hysteresis. The difference between the time profiles of the low energy intensity and high energy intensity (regressed to the low energy one) clearly demonstrates the magnetic cycle. If expressed as a hysteresis loop in the regression plot, the area of the loop is much greater for the even solar cycle.

In \cite{Krainev_etal_ICRC19_4_481-484_1985} we isolated two stages in the GCR intensity behavior in the maximum phase of solar cycle: 1) the first part of hysteresis connected with the postulated attenuation of $B^{HMF}$ during the HMF inversion  and not dependant on the charge $q$ of the GCR particle and the type of the inversion ($\propto dA/dt$) where $A$ is the dominant HMF polarity (the sign of $B_r^{HMF}$ in the north heliospheric hemisphere); and 2) the second part of hysteresis delayed with respect to the first stage by 1--2 years and dependant on the sign of $q\cdot dA/dt$. Now this scenario looks rather attractive for us if we connect the two above stages with two gaps in the double-gap structure of the GCR intensity and relate the postulated attenuation of $B^{HMF}$ with GG in this characteristic for the first stage while for the second stage substitute the sign of $dA/dt$ for the sign of $A$ itself (because of the "dipole" type of the HMF polarity distribution in this period, see the next section).

However, note that the magnitude of the discordance between the low and high energy GCR intensities during the first gap in the GCR intensity appears to be much greater for the even solar cycle 22 than for the odd SC 21 and 23 which looks as the dependence on $A$ or, rather, $dA/dt$ during the first stage. Certainly the choice of the regression period is very important and can change the features of the energy hysteresis. On the other hand the above feature can be due the fact that the reduction of $B^{HMF}$ is also much stronger for the even SC 22 than for the odd SC 21 and 23.

\section{What is the HMF inversion?}
If we discuss the GCR behavior during the inversion of the HMF polarity, we usually keep in mind the reversal of the radial components $B^{pol}_{ls}$ of the high--latitude large--scale SMF. However, the SMF does not directly influence the GCR intensity and to understand and model the GCR intensity during such a period one should have some model on what is going on with the HMF polarity distribution.

The main source of our notions on this distribution is the WSO model which can estimate $B_r^{ss}$ on the source surface $r_{ss}=(2.5\div 3.25)r_{Sun}$ in two variants of the inner boundary conditions: fixing from observations $B_{ls}^{ph}$ (classic) or $B_r^{ph}$ (radial) photospheric SMF components, see \cite{Hoeksema_PhD_1984} and references therein. So calculating $B_r^{ss}$ then finding isolines $B_r^{ss}=0$ and transporting this source surface neutral lines as the HCSs to the heliosphere by the solar wind, one can find the model of the HMF polarity distribution which is in a reasonable agreement with the observed crossings of the HCSs.
 \begin{figure}[!t]
  \centering
  \includegraphics[width=0.3\textwidth]{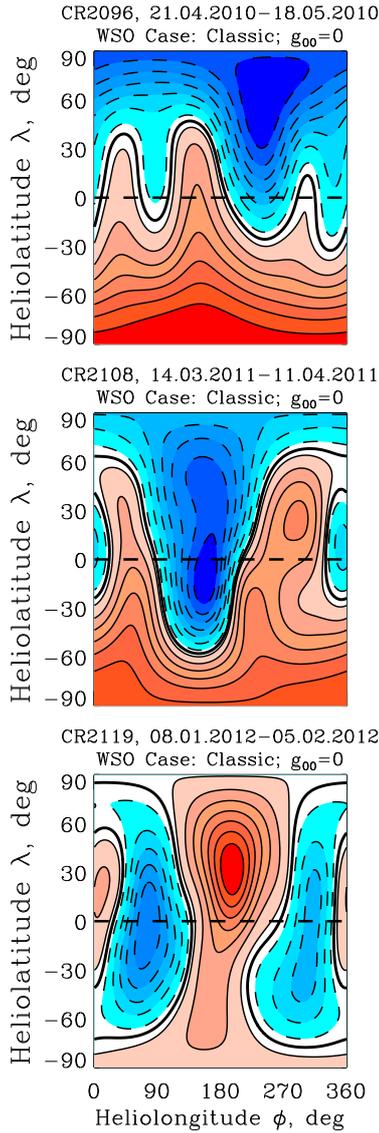}
  \caption{Three main types of the HMF polarity distribution. The thick solid lines are for the HCSs. The color (red for positive and blue for negative) stands for the HMF polarity while and its shades designate the magnitude of $B_r^{ss}$.}
  \label{fig2}
 \end{figure}

We calculated the above HMF polarity distribution for each Carrington rotations $N_{CR}=1642\div 2133$ (05.1976--02.2013) for two variants of the WSO model and both with and without the monopole term $g_{0,0}$ in the sets of the spherical harmonic coefficients. So the four cases of the HMF polarity distribution for each Carrington rotation were calculated. The corresponding classification of the time periods with respect to the HMF polarity distribution are shown as four horizontal lines in the upper part of Fig. \ref{fig1} (c) (classic--$g_{0,0}=0$, classic--$g_{0,0}\ne 0$,radial--$g_{0,0}=0$, radial--$g_{0,0}\ne 0$ from top to bottom).

All these distributions can be divided into three types different in the number and form of their HCSs and illustrated in Fig. \ref{fig2} for the case of the classic variant of WSO model without monopole term. The magnitude of $B_r^{ss}$ is not properly estimated in the WSO model, so correspondence between the magnitude and shades is not shown in Fig. \ref{fig2} and we discuss only the number and forms of their HCSs. The upper panel of Fig. \ref{fig2} illustrates the type of the HMF polarity distribution which we call the {\underline{"dipole"}} type characterized by the only and global HCS, that is HCS connecting all longitudes. This is the most common type of the HMF polarity distribution ($\approx 80\%$ of all time) both with $A<0$ (as in the upper panel of Fig. \ref{fig2}) and with $A>0$.
In the middle panel of Fig. \ref{fig2} another type of the HMF polarity distribution is shown which we call the {\underline{"transition dipole"}} type and which is also characterized by the global HCS but beside it one (as in the middle panel of Fig. \ref{fig2}) or several other HCSs exist. This type is less common ($\approx 10\%$) also both with $A<0$ (as in the middle panel of Fig. \ref{fig2}) and with $A>0$.
Finally we call the {\underline{"inversion"}} type the third type of the HMF polarity distribution illustrated in the lower panel of Fig. \ref{fig2}. It is characterized by the absence of the global HCS with several nonglobal HCSs (or even none at all which is common for the cases with the monopole spherical harmonic coefficient). This type is also much less common than the "dipole" type ($\approx 10\%$).

Beside this detailed classification of the HMF polarity distribution for each Carrington rotation we also consider the rough division of all time period into three corresponding types: 1) the "dipole" periods when all Carrington rotations are of the "dipole" type of the same dominating polarity $A$; 2) the "transition dipole"  periods limited by the Carrington rotations with this type of the HMF polarity distribution also of the same dominating polarity $A$; and 3) the "inversion" periods  limited by the Carrington rotations with this type. These types of periods are marked by the differently colored sections of the horizontal lines in the upper part of Fig. \ref{fig1} (c):  the "dipole" periods by the dark red (for $A>0$) or dark blue ($A<0$) sections;   the "transition dipole" periods by the light red ($A>0$) or light blue ($A<0$) sections; and the "inversion" periods are depicted by the black sections.

One can see in Fig. \ref{fig1} that the "dipole" period is characteristic for the low solar activity and its only HCS can be characterized by its waviness or quasi--tilt $\alpha_{qt}< \approx 40$ deg (for the classic variant of the WSO model). The "transition dipole" period is characteristic for the intermediate solar activity and it
corresponds to intermediate waviness of the HCS ($\approx 40<\alpha_{qt}< \approx 60$ deg.), but the quasi--tilt formally defined  as a half of the heliolatitude range of all HSCs is useless for the calculation of the drift magnetic velocity and the GCR intensity modeling. Finally, the "inversion" period is characteristic for the periods of maximum solar activity and the high--latitude SMF inversions and the formally defined quasi--tilt ($\alpha_{qt}> \approx 60$ deg) is also useless.

As can be seen in Fig. \ref{fig1} (c) for the fast and synchronous in the N-- and S--hemispheres SMF inversions (as in SC 21) the HMF inversion period is also short, while for the prolonged and nonsynchronous SMF inversions (as in SC 22 and 23) the HMF inversion periods are also longer. In general the HMF inversion periods are centered with their SMF counterparts and approximately coincide with the Gnevyshev Gap in the sunspot area and HMF strength. Usually the HMF inversion periods are surrounded by comparable--sized "transition dipole" periods. During the main part of solar cycle the HMF polarity distribution is "dipole"--like. This dipole period is asymmetrical with respect to the moment $t^{GCR}_{max}$ of the maximum of the GCR intensity, (approximately $t^{GCR}_{max}-6<t$, years $<t^{GCR}_{max}+1.5$). Note that the second gap and the fast or slow  increase in the GCR intensity occur in the beginning of this period, more than 4 or 5 years before $t^{GCR}_{max}$. This fact is important as the magnetic drift is usually considered significant only in the periods around solar minima \cite{Jokipii_Wibberentz_SSR_83_365_1998}.

So for the phase of the low sunspot activity with the "dipole" type of the HMF polarity distribution the GCR intensity can be calculated using the transport equation with the usual magnetic drift velocity terms (e. g., utilizing the tilted-CS model with a tilt $\alpha_t$ as a parameter) and getting $\alpha_t$ as the quasi--tilt $\alpha_{qt}$ from \cite{WSO_Site}. However, how to get these terms for the high sunspot activity phase with the "transition dipole" and  "inversion" types of the HMF polarity distributions, when there are several (or none) HCSs and the formally defined quasi--tilt is useless? As in \cite{KalininKrainev_ECRS21_222-225_2009} the regular 3D HMF can be represented as $\vec{\cal B}(r,\vartheta,\varphi)={\cal F}(r,\vartheta,\varphi)\vec{\cal B}^m(r,\vartheta,\varphi)$,
where $\vec{\cal B}^m$ is the unipolar
(or ``monopolar'') magnetic field and the HMF polarity ${{\cal F}}$ is a scalar function equal to $+1$ in the
positive and $-1$ in negative sectors,
changing on the HCS surface ${{\cal F}} (r,\vartheta,\varphi,t)=0$.
Then the 3D particle drift velocity is
${\vec{{\cal V}}}^d=pv/3q\left[{\bf{\nabla}}\times(\vec{{\cal B}}/{{\cal B}}^2)\right]$, \cite{RossiOlbert_1970},
where $v$ and $q$ are the particle speed and
charge, respectively. One can decompose the drift velocity into the regular and current sheet velocities:
\begin{eqnarray}
{\vec{{\cal V}}}^{d,reg}=pv/3q{{\cal F}}\left[{\bf{\nabla}}\times(\vec{{\cal B}}^m/{{\cal B}}^2)\right]\\
{\vec{{\cal V}}}^{d,cs}=pv/3q\left[{\bf{\nabla}}{{\cal F}}\times(\vec{{\cal B}}^m/{{\cal B}}^2)\right].
\end{eqnarray}

So to get the magnetic drift velocities for any type of the HMF polarity distribution one needs only ${\cal F}$ and ${\bf{\nabla}}{\cal F}$ or in 2D case $F$ and $dF/d\vartheta$, where $F$ is the HMF polarity ${\cal F}$ averaged over the longitude. All of these quantities (${{\cal F}},\nabla{\cal F},F,dF/d\vartheta$) can be calculated numerically for any calculated HMF polarity distribution. Note that the fundamental difference between the global and nonglobal HCS is in the fact that the sign of the radial component of the current sheet drift changes as the particle moves along the  nonglobal HCS, so that the connection between the inner and outer heliosphere is blocked.

\section{On the maximum phase of SC 24}
For the current SC 24 the "dipole" period ended in the beginning of 2011. For the classic cases the "transition dipole" period ended in the beginning of 2012 while for the radial cases there is no such period. So in the current solar cycle the HMF inversion started well before both the SMF inversion and the first peak in the $S_{ss}$ and $B_{HMF}$ (02.2012; the only peak of the double--peak structure). Neither gaps nor energy hysteresis are observed in the GCR intensity up to now. In the beginning of 2013 we are still in the middle of the HMF inversion.
This unusual features of the current HMF inversion are probably connected with the unusually low solar and heliospheric activity in the last two solar cycles.

As one can see from  panels (a) and (b) of Fig. \ref{fig1} both the sunspot and high--latitude solar activity and the HMF strength are very low during the ascending phase of SC 23 and the minimum 23/24 between SC 23 and 24. As a result the GCR intensity in the minimum 23/24 is the highest ever measured (see \cite{Stozhkov_etal_Astrophysics_SpaceSciences_Transactions_7(3)_379-382_2011,Mewaldt_etal_AstrophysJLetters_723_2010,Gushchina_etal_Journal_of_Physics_Conference_Series_409_012169_2013})
and its current value is still much higher than in the maxima of the previous solar cycles.

However, in the documented history of the solar characteristics there were long periods of the high (global maxima) and low (global minima) sunspot activity  \cite{Schove_1983}, and the cycles of the second half of the last century belong to so called Modern maximum. As we demonstrated in \cite{Krainev_Kalinin_Journal_of_Physics_Conference_Series_409_012176_2013} up to now the sunspot area in SC 24 is much lower than it was in the cycles of the Modern maximum but much greater than in the Maunder and even Dalton minima. In general it corresponds to the Glaisberg minimum in the first decades of the last century. Using the regression between the values of both characteristics in the first regression points and in the maximum of the previous solar cycles we managed to estimate the maximum $S_{ss}^{M24}$ and minimum $J^{M24}$ expected for the maximum of SC 24 and came to the conclusion that in SC 24 the maximum sunspot area can be of the highest values for the Glaisberg minimum (SC 14--16). As to the GCR intensity, our upper estimate of $J^{M24}$ indicates that in SC 24 the minimum GCR intensity can be slightly higher than in SC 20, 21, 23.

As we stated in \cite{Krainev_Kalinin_Journal_of_Physics_Conference_Series_409_012176_2013} up to 2012 the development of SC 24 in the N--hemisphere was like in the "Modern Maximum" (SC 17--23), while that in the S--hemisphere more closely resembled the Dalton minimum (SC 5--7). During the last year (2012) the sunspot area in the N--hemisphere decreased while that in the S--hemisphere increased and if the sunspot activity in the S--hemisphere overtakes that in the N--hemisphere during the maximum phase of solar cycle (as is often the case) then a lot depends on the activity in the south solar hemisphere in the nearest future.

\section{Conclusions}
\noindent 1. The two--stage scenario of the  main characteristic features of the GCR intensity behavior in the maximum phase of solar cycle is suggested: 1) the first gap of double--gap structure and the first part of hysteresis occur during the inversion of the heliospheric magnetic field while 2) the second gap and the second part of hysteresis proceeds during the periods characterized by the "dipole" type of the HMF polarity distribution leading to the magnetic cycle in the GCR intensity.

\noindent 2. We isolate three main types of the HMF polarity distribution: 1) the "dipole" type with the only and global HCS characteristic for the periods of the low solar activity; 2) the "transition dipole" type with the global HCS and several other HCSs characteristic for the periods of the intermediate solar activity and 3) the "inversion" type with the absence of the global HCS  characteristic for the periods of HMF inversion. The simple procedure to get the magnetic drift velocities from the calculated HMF polarity distribution is discussed.

\noindent 3. The comparison of the current solar cycle 24 in the sunspot activity and GCR intensity with the past solar cycles  shows that sunspot activity corresponds to the Glaisberg minimum in the first decades of the last century while the GCR intensity is slightly higher than in previous solar cycles. The estimation is made of the sunspot area and GCR intensity  expected for the maximum of SC 24. The nearest future of the SC 24 depends on the activity in the south solar hemisphere.

\vspace*{0.5cm}
\footnotesize{{\bf Acknowledgment:}{ We thank the Russian Foundation for Basic Research (grants 11-02-00095a, 12-02-00215a, 13-02-00585a, 13-02-10006k) and the Program "Fundamental Properties of Matter and Astrophysics" of the Presidium of the Russian Academy of Sciences.}}

\end{document}